\documentclass[twocolumn,twoside,slac_two]{revtex4}
\usepackage{graphicx}
\usepackage{fancyhdr}
\def\units#1{\hbox{$\,{\rm #1}$}}
\def\degrees{\hbox{$^\circ$}}

\begin{document}

\title{Unfolding spectral analysis of the Fermi-LAT data}

%

\author{F.~Loparco}
\affiliation{Universit\`a di Bari and INFN Sezione di Bari}
\author{M.~N.~Mazziotta}
\affiliation{INFN Sezione di Bari}

\author{(on behalf of the Fermi LAT Collaboration)}

\begin{abstract}

The Large Area Telescope (LAT) onboard the {\em Fermi} 
satellite is observing the gamma-ray sky in the high 
energy region, above $20 \units{MeV}$. We have developed a 
method to reconstruct the energy spectra of the gamma-rays 
detected by the Fermi LAT instrument based on a Bayesian 
unfolding approach, that takes into account the energy dispersion
introduced by the instrument response. The method has been
successfully applied to reconstruct the energy spectra of
both steady and pulsating point sources. The analysis
technique will be illustrated and the results obtained in
some significant test cases will be discussed.

\end{abstract}

\maketitle

\thispagestyle{fancy}


\section{Introduction}

One of the main tasks of physics data analysis is to 
reconstruct the true distribution of a given physical 
quantity from the observed one, getting rid of the 
distorsions introduced by the detector response and
of the background events. This can be accomplished by
different methods, that follow different approaches.

A possible strategy is to assume that the true distribution 
(and the noise too) is described by a mathematical function
depending on a set of free parameters, that have to be
estimated. This approach is called parametric inference
(fits) and is implemented in the well known least squares
or maximum likelihood methods. Parametric inference is 
a valid choice if there are good reasons to believe that
the true distribution can be described by a given analytical 
function.

On the other hand, sometimes physicists don't need to
interpret the observed data in the framework of a model,
but only wish to estimate the true distribution with its
uncertainties. This second approach goes under the name
of unfolding (or deconvolution). 

In this paper we will illustrate an unfolding method to 
reconstruct the energy spectra of point like gamma-ray
sources detected by the Large Area Telescope 
(LAT)~\cite{Atwood2009} operating on the {\em Fermi} 
satellite. The applications of the method to the cases 
of an isolated point source and of a pulsating source 
will be also discussed.

\section{The unfolding method}

The unfolding method that we have developed allows
to reconstruct the true energy spectra of gamma-ray
point sources detected by the {\em Fermi} LAT from the
observed ones, taking into account the instrument
response function (IRF). The procedure can be applied
to both steady and pulsating point sources.

\subsection{Event selection}
\label{sec:evesel}

For our analysis we usually 
select photon events within an energy 
dependent region of interest (ROI) centered on the source 
position. The maximum allowed angular separation from the 
source is given by:
\begin{equation}
\label{eq:psf}
\theta_{max} = max \{ min[5\degrees, 
5\degrees (100\units{MeV}/E)^{0.8}] , 0.35\degrees \}
\end{equation}
where $E$ is the photon energy. The function in 
eq.~\ref{eq:psf} reproduces the energy dependence of 
the LAT $68\%$ angle containment up to a few 
$\units{GeV}$ for photons converting in the back 
section of the instrument. 
The size and the shape of the
analysis ROI can be changed by the user,
according to the kind of source to be 
investigated. 

Photon entering in the detector with zenith angles
larger than $66.4\degrees$ in the instrument frame
are excluded from the analysis.
Photons with zenith angles larger than $105\degrees$
with respect to the Earth reference frame are
also disregarded in order to avoid contamination
from Earth albedo.

\subsection{Background subtraction}

A crucial point for the analysis (that is not 
connected with the method!) is the evaluation of
background, that has to be subtracted from the observed 
energy spectra. The background includes an isotropic
component, a galactic diffuse component and all the
contributions from nearby sources. Its energy spectrum 
can be evaluated either from real data or from a model. 

In the first case the background events are evaluated 
selecting the photons coming from an annulus external to 
the ROI used for the analysis. The background energy 
spectrum built in this way is then properly rescaled
taking into account the solid angle and observation
time ratios. This approach is valid if the background
distribution does not exhibit significant spatial
gradients, as in the case of point sources far from
the galactic plane. When this condition does not
occur, as in the case of point sources close to the
galactic planes, the background observed spectrum must 
be evaluated using a proper background model.

The background spectra can be safely evaluated from 
real data also when a pulsating point source is being
investigated. Usually, for this class of sources, it
is interesting to study only the pulsed emission.
In this case the timing information allows to build
the observed spectrum selecting photons within the ROI 
and with phases in the on-pulse range, and the background 
spectrum selecting photons within the ROI and with
phases in the off-pulse range. The background subtraction
is then performed taking the on/off pulse ratio into
account.
  
\subsection{The smearing matrix}
\label{sec:sm}

The relation between the observed energy spectrum 
is given by the following equation:
\begin{equation}
N(E_{obs,i}) = \sum_{j} P(E_{obs,i} | E_{true,j}) N(E_{true,j})
\end{equation}
where $N(E_{obs,i})$ is the number of events observed
in the $i$-th observed energy bin, $N(E_{true,j})$ is the
number of real events in the $j$-th true energy bin 
and $P(E_{obs,i} | E_{true,j})$ is the probability that 
a photon with true energy $E_{true,j}$ is detected
with an observed energy $E_{obs,i}$. The probabilities
$P(E_{obs,i} | E_{true,j})$ are the elements of the so called
smearing matrix, which condenses the information 
about the IRF.

It is worth to point out that the smearing matrix
includes efficiencies and acceptance. In fact, the sum
\begin{equation}
\epsilon_{j} = \sum_{i} P(E_{obs,i} | E_{true,j}) < 1 
\label{eq:eff}
\end{equation}
represents the probability that a photon generated
at the source with true energy $E_{true,j}$ is detected
with any value of observed energy. 

The smearing matrix is evaluated using a Monte Carlo
simulation performed with Gleam, the full LAT simulation 
package based on the Geant4 toolkit~\cite{Mazziotta2009}. 
A trial photon spectrum is
simulated and the elements of the smearing matrix are
evaluated as:
\begin{equation}
P(E_{obs,i} | E_{true,j}) =  \frac{N(E_{obs,i} | E_{true,j}, cuts)}{N(E_{true,j})}
\end{equation} 
where $N(E_{true,j})$ are the events generated in the $j$-th 
true energy bin and $N(E_{obs,i} | E_{true,j}, cuts)$ are the events 
generated in the $j$-th true energy bin and detected in the
$i$-th observed energy bin after applying the analysis cuts.
The incoming photon directions are simulated according to
the source pointing history.

\subsection{Unfolding analysis}

Once the observed spectrum and the smearing matrix are
evaluated, the source spectrum is evaluated by means of
a Bayesian unfolding iterative procedure~\cite{DAgostini1995}.
Each iteration goes through the following steps:
\begin{enumerate}
\item{the spectrum $N(E_{true,j})$ reconstructed after the 
previous iteration is assumed as starting point;}
\item{the smearing matrix is inverted by applying Bayes'theorem:
\begin{equation}
P(E_{true,j} | E_{obs,i}) = \frac{P(E_{obs,i} | E_{true,j}) P(E_{true,j})}
{P(E_{obs,i})}
\end{equation}
where $P(E_{obs,i})$ and $P(E_{true,j})$ are the normalized
observed and true spectra;}
\item{a new true spectrum is evaluated using the inverted
smearing matrix:
\begin{equation}
N(E_{true,j}) = 
\frac{1}{\epsilon_{j}} \sum_{j} P(E_{true,j} | E_{obs,i}) N(E_{obs,i}).
\label{eq:unfo}
\end{equation}
}
\end{enumerate}

As a starting point of the first iteration a trial spectrum (usually
a flat one) is assumed. The iterative procedure is terminated when 
the spectrum reconstructed after the last iteration does not differ
significantly from the previous one (a $\chi^{2}$ test is performed).
Usually the convergence is reached after a few iterations and the
results does not depend on the input trial spectrum.

After the unfolding procedure has converged, it is possible
to calculate the errors associated to the unfolded points by
applying error propagation to eq.~\ref{eq:unfo}~\cite{DAgostini1995}.
Since the unfolded points are correlated, the errors must
be described by means of a covariance matrix. We have 
implemented in our code a procedure to evaluate the contributions
to the error matrix from both statistical and systematic errors. 
For what concerns statistical errors, we have taken into
account the contribution from the limited statistics of the
observed data sample as well as the contribution from the
finite statistics of the Monte Carlo sample used to evaluate
the smearing matrix. Systematic errors arise from the 
limited knowledge of the IRF, and are evaluated assuming
an energy uncertainty depending logarithmically on the
observed energy, that changes from $10\%$ below $100\units{MeV}$
to $5\%$ at $562\units{MeV}$ and to $20\%$ above $10\units{GeV}$. 
The full covariance matrix among unfolded points must be used
when the unfolded flux has to be fitted with some parametric
function~\cite{Mazziotta2009}.

\section{Application to simulated data sets}

In this section the results obtained applying the unfolding procedure
to simulated data sets will be discussed. Simulations have been 
performed using the {\em Fermi} simulation toolkit {\em gtobssim}.

\subsection{Application to a steady point source}

\begin{figure}[ht]
\includegraphics[width=75mm]{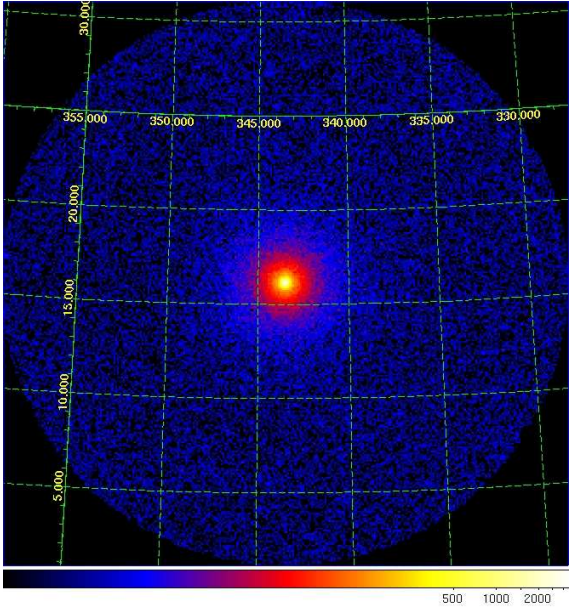}
\caption{Photon count map in a $15\degrees$ region around the 3C454.3
simulated source.}
\label{fig:map3C}
\end{figure}

\begin{figure}[h]
\includegraphics[width=75mm]{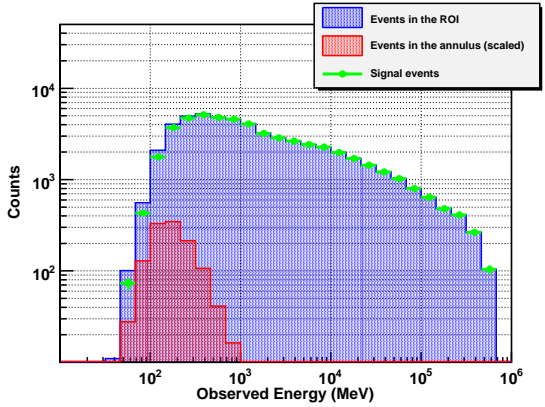}
\caption{Observed energy spectra for the 3C454.3
simulated point source. The background
counts, that are obtained considering photons 
in an annulus with an inner radius of $14\degrees$ 
and an outer radius of $15\degrees$, are scaled for
the solid angle and live time ratios, and are then
subtracted from the counts in the analysis ROI.}
\label{fig:counts3C}
\end{figure}

A steady point source has been simulated in the 3C454.3 sky
position $(343.52\degrees, 16.16\degrees)$. The source has been
simulated for a total time of $200\units{days}$, assuming a 
power law spectrum given by:
\begin{equation}
\frac{dN}{dE} = k \left( \frac{E}{E_{0}} \right)^{-\Gamma}
\label{eq:spl}
\end{equation}
where $E_{0}=1\units{GeV}$ is the scale energy. 
The spectral index value
is $\Gamma=1.5$ and the value of $k$ has been chosen to 
have a flux above $100\units{MeV}$ of 
$6\times10^{-6} \units{photons~cm^{-2}~s^{-1}}$. Both galactic and
extragalactic background components have been included in 
the simulation and the true LAT pointing history has been
used.

\begin{figure}[ht]
\includegraphics[width=75mm]{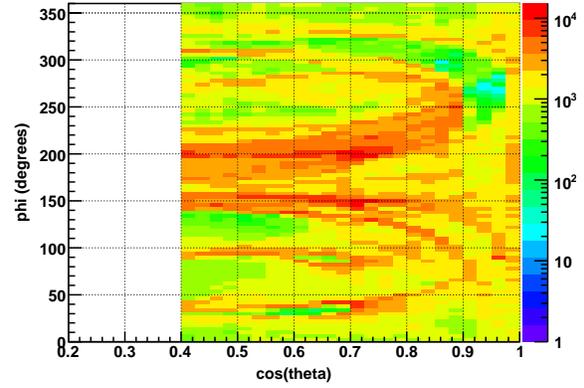}
\caption{Angular distribution of the photons expected
from the 3C454.3 simulated source.}
\label{fig:pointing3C}
\end{figure}

\begin{figure}[h]
\includegraphics[width=75mm]{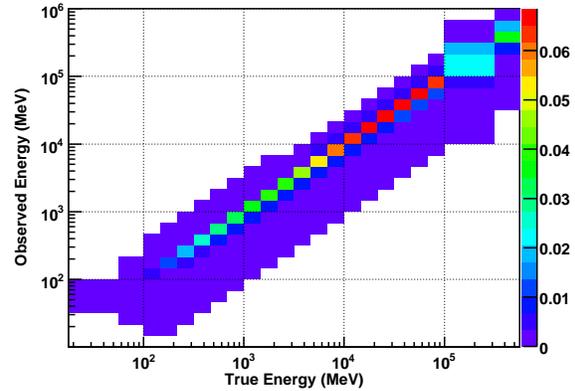}
\includegraphics[width=75mm]{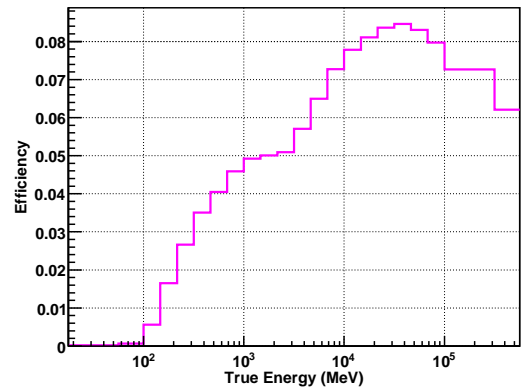}
\caption{Smearing matrix (a) and efficiency (b) 
for the 3C454.3 simulated point source.}
\label{fig:eff3C}
\end{figure}

\begin{figure}[ht]
\includegraphics[width=75mm]{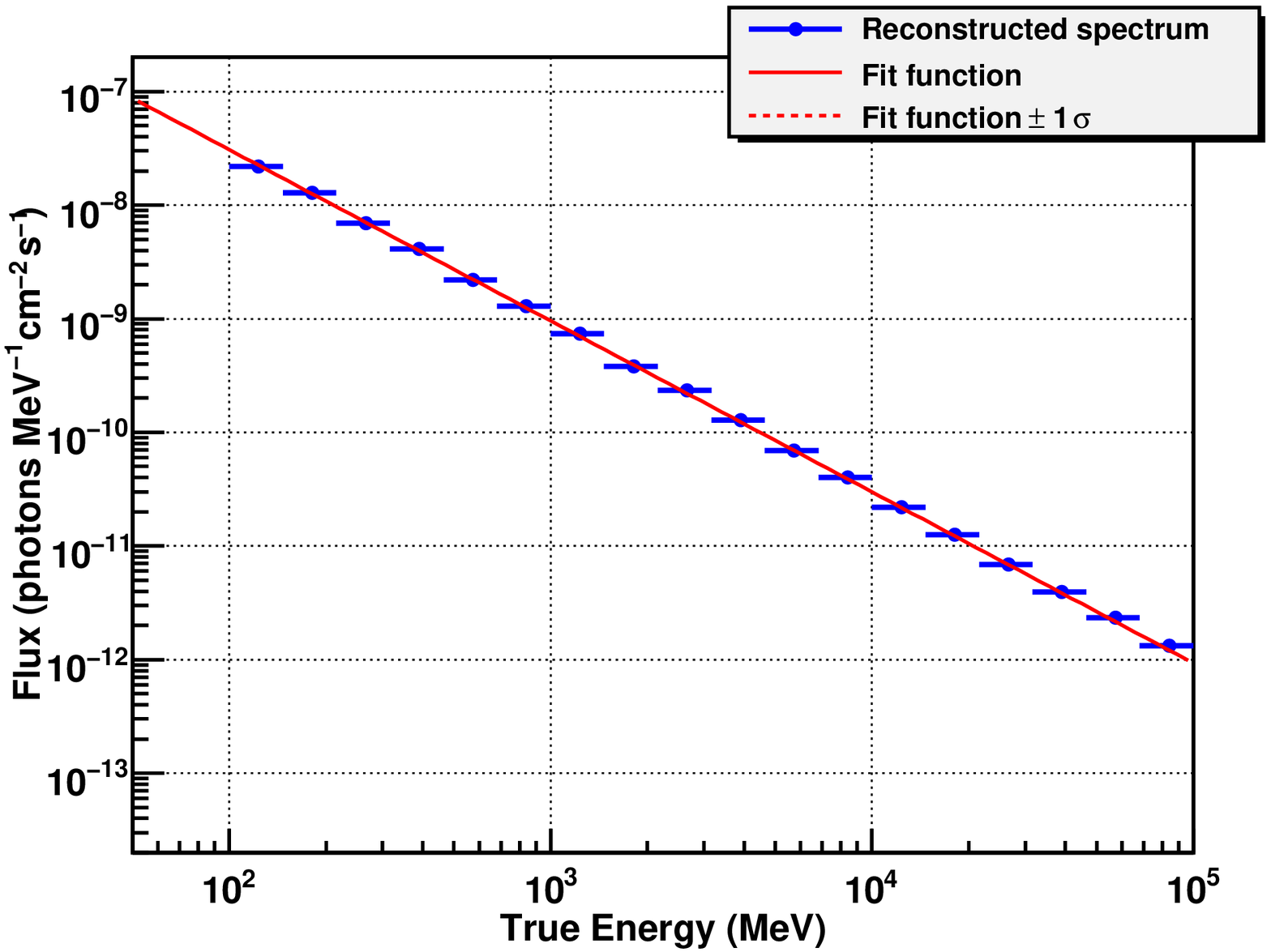}
\includegraphics[width=75mm]{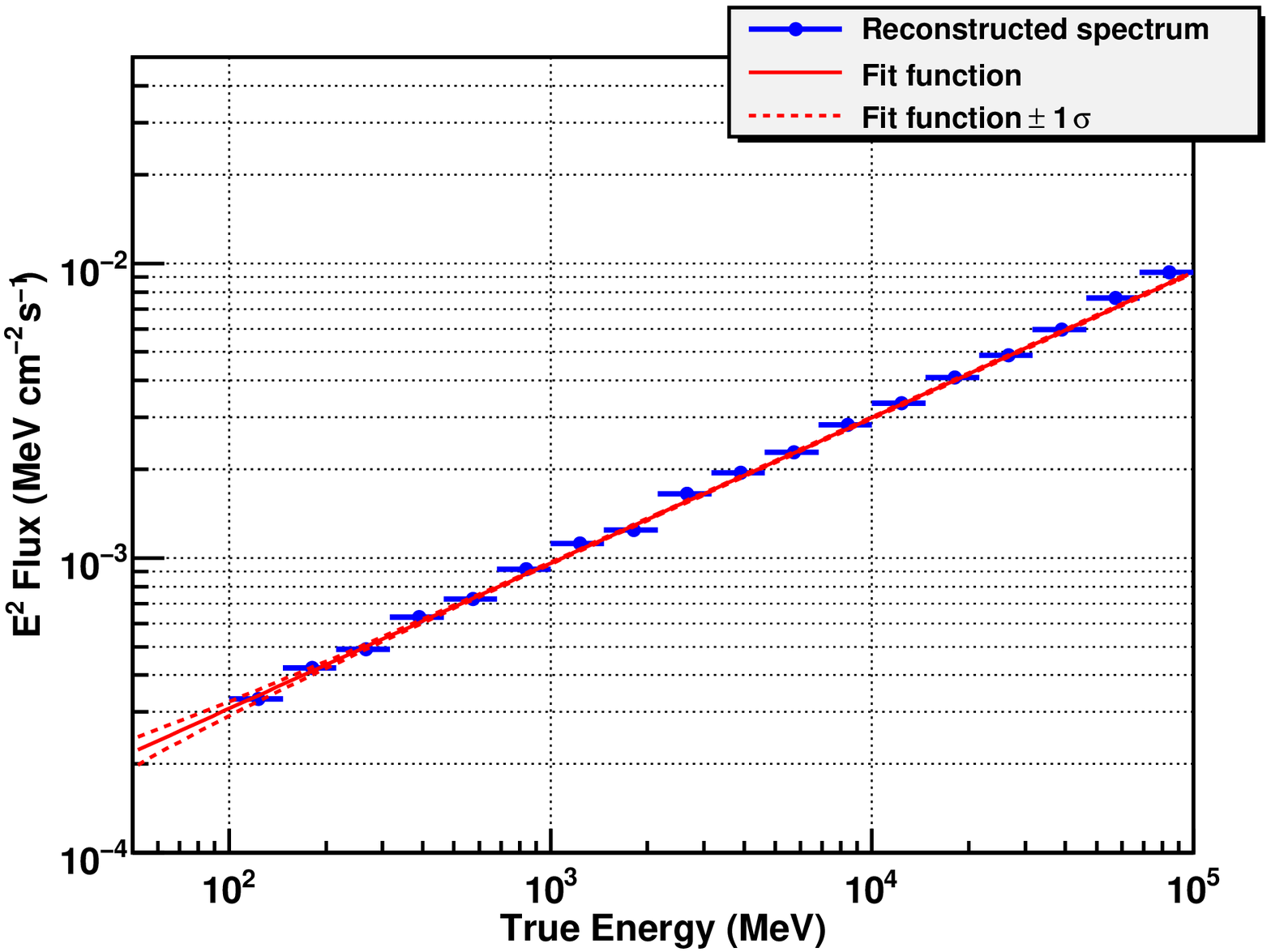}
\caption{Reconstructed energy spectrum and SED
of the simulated 3C454.3 point source. The
spectrum has been fitted with a simple power
law. The fit function together with the
$\pm 1 \sigma$ error regions are also shown.}
\label{fig:unfo3C}
\end{figure}

Fig.~\ref{fig:map3C} shows the photon count map in a
$15\degrees$ region centered on the position of the simulated 
source. Since the source is far away from the galactic plane,
the background is nearly isotropic and can be safely
evaluated from real data. The analysis of the 3C454.3 
point source has been performed selecting photons 
in the ROI given by eq.~\ref{eq:psf} and evaluating 
the background in an annulus with an inner radius 
of $14\degrees$ and an outer radius of $15\degrees$. 
In fig.~\ref{fig:counts3C} the observed energy spectrum 
is shown together with the background energy
spectrum. The signal spectrum is obtained after
subtracting the background spectrum from the
observed one, after proper normalization.

In fig.~\ref{fig:pointing3C} it is shown the
expected angular distribution in the instrument
coordinates ($cos \theta$ and $\phi$) of the
photons expeted from the 3C454.3 simulated source.
The distribution has been built using the 
information from the spacecraft pointing 
history. As it is evident from the figure, 
the instrument coordinate phase space is not
uniformly populated. This behaviour is due to
the relative motion of the satellite with 
respect to the source, and is taken into account
when calculating the smearing matrix.
The region $cos\theta < 0.4$ is empty because of
the angular selection cut illustrated in
section~\ref{sec:evesel}.

\begin{figure}[ht]
\includegraphics[width=75mm]{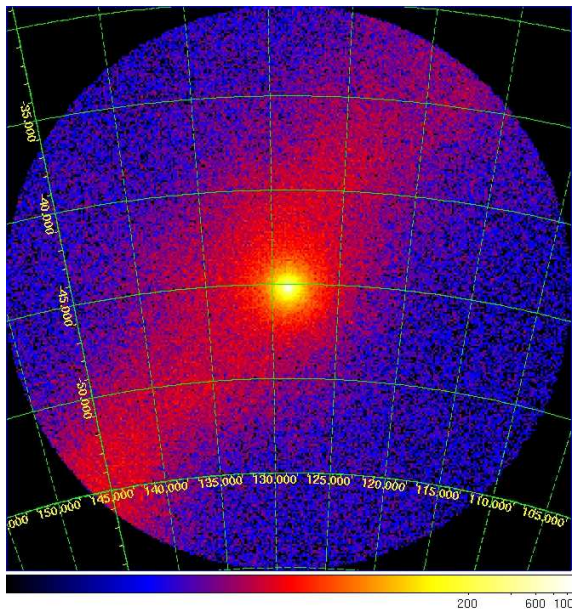}
\caption{Photon count map in a $15\degrees$ region around the 
Vela simulated source.}
\label{fig:mapvela}
\end{figure}

Fig.~\ref{fig:eff3C}.a shows the smearing matrix
evaluated for the simulated 3C454.3 point source,
using the procedure described in section~\ref{sec:sm}.
In fig.~\ref{fig:eff3C} the effect of energy dispersion
is clearly evident. If there were no energy dispersion,
the smearing matrix would be diagonal; the presence
of energy dispersion implies off-diagonal terms
in the smearing matrix.   
In fig.~\ref{fig:eff3C}.b it is shown the efficiency
as a function of the true energy, evaluated 
according eq.~\ref{eq:eff}. The efficiency increases
with energy reaching a maximum of about $8.5\%$ at
a few tens of $\units{GeV}$ and drops at $6\%$ at
larger energies.

Fig.~\ref{fig:unfo3C} shows the energy spectrum 
of the simulated 3C454.3 source reconstructed with
the unfolding procedure. The meaured flux above $100\units{MeV}$ 
is of $(5.94 \pm 0.08) \times 10^{-6} \units{photons~cm^{-2}~s^{-1}}$
and is consistent with the true value.
The unfolded spectrum has been also fitted with 
the function of eq.~\ref{eq:spl} using the covariance
matrix provided by the spectral reconstruction algorithm. 
The fit parameters are $\Gamma=1.51 \pm 0.01$ and 
$k=(9.60 \pm 0.02)\times 10^{-10} \units{photons cm^{-2}~s^{-1}}$,
corresponding to a flux above $100\units{MeV}$ of
$(6.08 \pm 0.12) \times 10^{-6} \units{photons~cm^{-2}~s^{-1}}$,
that is still consistent with the input value.

\subsection{Application to a pulsar}

A pulsar has been simulated in the Vela sky 
position $(128.83\degrees, -45.18\degrees)$. The source
has been simulated for $200\units{days}$, assuming a 
pulsed spectrum given by:
\begin{equation}
\frac{dN}{dE} = k \left( \frac{E}{E_{0}} \right)^{-\Gamma}
\exp \left(- \frac{E}{E_{cut}} \right)
\label{eq:ec}
\end{equation}
where $E_{0}=1\units{GeV}$ is the scale energy.
The spectral index value is $\Gamma=1.62$, the 
cutoff energy is $E_{cut}=8\units{GeV}$ and the value of $k$
has been chosen to have a flux above  $100\units{MeV}$ 
of $8.5\times10^{-6} \units{photons~cm^{-2}~s^{-1}}$. The scale
energy $E_{0}$ has been set to $1\units{GeV}$.
Galactic and extragalactic background components 
have been included in the simulation and the true 
LAT pointing history has been used. A custom 
light curve has been used to simulate pulsed emission
for a $88\%$ fraction of the total phase interval.

\begin{figure}[ht]
\includegraphics[width=75mm]{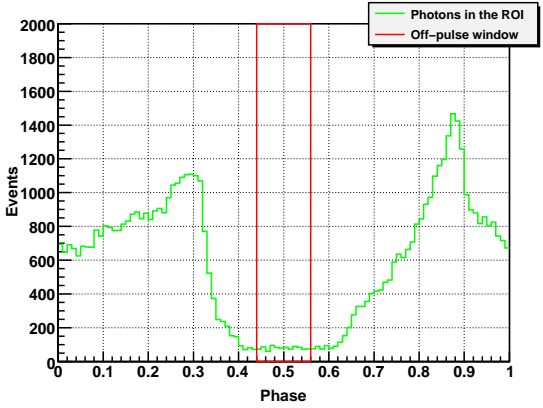}
\caption{Phase distribution of photons in the analysis ROI.
The off-pulse window is defined as the phase interval 
from $0.44$ to $0.56$.}
\label{fig:lightcurve}
\end{figure}

\begin{figure}[h]
\includegraphics[width=75mm]{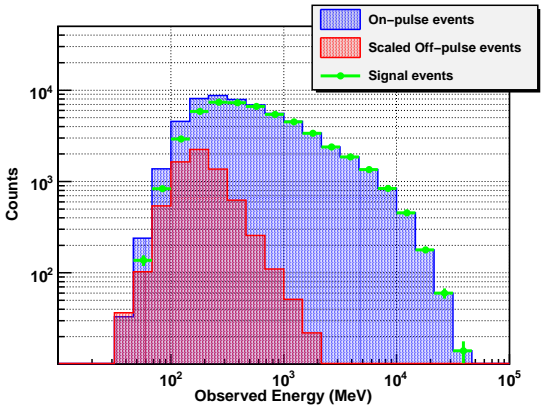}
\caption{Observed energy spectra for the simulated
Vela point source. The background counts, 
that are obtained considering off-pulse photons 
in the analysis ROI, are scaled for
phase ratio $0.88/0.12$, and are then
subtracted from the on-pulse counts.}
\label{fig:countsvela}
\end{figure}

Fig.~\ref{fig:mapvela} shows the photon count map in a
$15\degrees$ region centered on the position of the simulated 
source. In this case the source lies on the galactic 
plane and the background is not isotropic. However,
since we are intersted in the study of the pulsed
component of the spectrum, this is not an issue.
In fig.~\ref{fig:lightcurve} it is shown the
light curve obtained selecting photons in the analysis
ROI of eq.~\ref{eq:psf}. The off-pulse window is defined 
as the phase interval from $0.44$ to $0.56$.

The analysis of the Vela point source has been 
performed selecting photons in the ROI given by 
eq.~\ref{eq:psf}, with arrival times in the on-pulse 
window. Fig.~\ref{fig:countsvela} shows the
observed spectrum together with the background 
spectrum, that is obtained selecting off-pulse photons.
The signal spectrum is evaluated subtracting the 
off-pulse counts rescaled for the phase ratio 
$0.88/0.12$ to the on-pulse ones. 

\begin{figure}[ht]
\includegraphics[width=75mm]{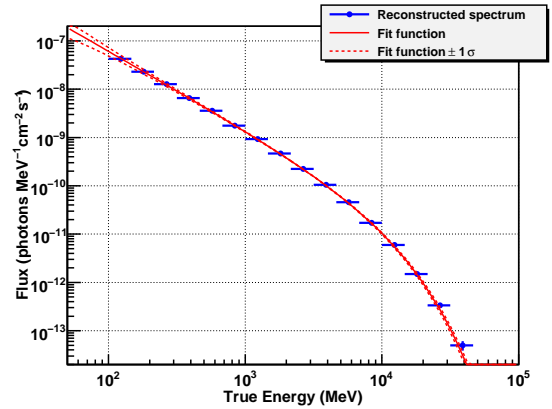}
\includegraphics[width=75mm]{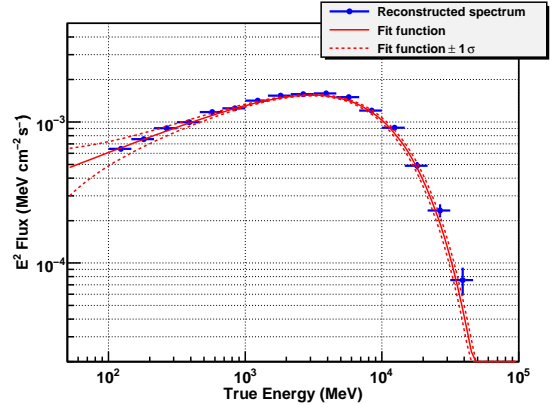}
\caption{Reconstructed pulsed energy spectrum and SED
of the simulated Vela point source. The
spectrum has been fitted with a power
law with exponential cutoff. The fit function 
together with the $\pm 1 \sigma$ error regions 
are also shown.}
\label{fig:unfovela}
\end{figure}

Fig.~\ref{fig:unfovela} shows the pulsed energy spectrum 
of the simulated Vela source reconstructed with
the unfolding procedure. The meaured flux above $100\units{MeV}$ 
is of $(8.49 \pm 0.08) \times 10^{-6} \units{photons~cm^{-2}~s^{-1}}$
and is in agreement with the true value.
The unfolded spectrum has been also fitted with 
the function of eq.~\ref{eq:ec}. The fit parameters
are $\Gamma=1.62 \pm 0.02$, $E_{cut}=(8050\pm330)\units{MeV}$ and 
$k=(1.48 \pm 0.02)\times 10^{-11} \units{photons cm^{-2}~s^{-1}}$,
corresponding to a flux above $100\units{MeV}$ of
$(8.60 \pm 0.15) \times 10^{-6} \units{photons~cm^{-2}~s^{-1}}$,
that is consistent with the input value.

\section{Conclusions}

We have developed a method to reconstruct the energy 
spectra of both steady and pulsating point sources
detected by the {\em Fermi} LAT, based on a bayesian
unfolding technique. This analysis technique, that
takes into account the energy dispersion introduced
by the detector and allows to reconstruct the spectra
without the need of assuming any parametric model,
has been successfully tested with simulated data sets.

If an unfolded spectrum needs to be interpreted in the
framework of a given model, it can be easily fitted with
any function, taking into account the full covariance
matrix, that is evaluated at the end of the unfolding
procedure, and that includes both statistical and
systematic uncertainties.


\bigskip 
\begin{acknowledgments}

The {\em Fermi} LAT Collaboration acknowledges support from
a number of agencies and institutes for both development and
the operation of the LAT as well as scientific data
analysis. These include NASA and DOE in the United States,
CEA/Irfu and IN2P3/CNRS in France, ASI and INFN in
Italy, MEXT, KEK and JAXA in Japan, and the 
K.~A.~Wallenberg Foundation, the Swedish Research
Council and the National Space Board in Sweden.
Additional support from INAF in Italy for science
analysis during the operation phase is also
gratefully acknowledged.

\end{acknowledgments}

\bigskip 


\begin{thebibliography}{9}   

\bibitem{Atwood2009}
W.~B.~Atwood et al., \emph{Astrophys. J.} \textbf{697}, 1071 (2009).

\bibitem{Mazziotta2009}
M.~N.~Mazziotta, \emph{Proc. of the 31st ICRC, Lodz 2009},
arXiv:0912.1236

\bibitem{DAgostini1995}
G.~D'Agostini, \emph{Nucl. Instr. Meth.} \textbf{A 362}, 487 (1995).


\end{thebibliography}

\end{document}